\documentclass[a4paper]{jpconf}

\usepackage{setspace}
\usepackage{graphicx}

\newlength{\figwidth}
\newlength{\fighwidth}
\def\P{\vec P}

\newcommand{\be}{\begin{equation}}
\newcommand{\ee}{\end{equation}}
\newcommand{\bge}{\begin{equation}}
\newcommand{\ene}{\end{equation}}
\newcommand{\bea}{\begin{eqnarray}}
\newcommand{\eea}{\end{eqnarray}}
\newcommand{\bg}{\begin{eqnarray}}
\newcommand{\en}{\end{eqnarray}}

\begin{document}

\title{Medium Modifications of Hadron Properties and Partonic Processes}

\author{W. K. Brooks$^1$, S. Strauch$^2$ and K. Tsushima$^3$}
\address{
$^1$ Universidad T\'ecnica Federico Santa Mar\'ia, Valpara\'iso, Chile\\
$^2$ University of South Carolina, Columbia, South Carolina 29208, USA	\\
$^3$ University of Adelaide, South Australia 5005, Australia}

\begin{abstract}
Chiral symmetry is one of the most fundamental symmetries in QCD. It is closely connected to hadron properties in the nuclear medium {\it via} the reduction  
of the quark condensate $<\bar{q}q>$, manifesting the partial restoration of chiral symmetry. 
To better understand this important issue, a number of Jefferson Lab experiments over the past decade  have focused on understanding properties of mesons and nucleons 
in the nuclear medium, often benefiting from the high polarization and 
luminosity of the CEBAF accelerator. In particular, a novel, accurate, polarization transfer measurement technique 
revealed for the first time a strong indication that the bound proton electromagnetic
form factors in $^4$He may be modified compared to those in the vacuum.
Second, the photoproduction of vector mesons on various nuclei has been
measured {\it via} their decay to $e^+e^-$ to study possible in-medium
effects on the properties of the $\rho$ meson. In this experiment, no
significant mass shift and some broadening consistent with expected
collisional broadening for the $\rho$ meson has been observed,
providing tight constraints on model calculations.
Finally, processes involving in-medium parton propagation have been studied.
The medium modifications of the quark fragmentation functions have been
extracted with much higher statistical accuracy than previously possible.
\end{abstract}

\section{Introduction}

One of the most exciting topics in nuclear physics is to study how the
properties of hadrons are modified by the nuclear environment,
and how such modifications are related to the properties of nuclei and nuclear phenomena.
Since nucleons and mesons are made of quarks, antiquarks and gluons, one expects their
internal structure to be modified when placed in a nuclear medium or
atomic nuclei~\cite{Brown91}. From the QCD point of view, this is primarily the manifestation of  modifications of the quark condensate in the nuclear medium, which is
closely associated with the partial restoration of chiral symmetry. Similarly, medium modifications of partonic processes are directly tied to the gluon density of the medium. The modern view of the nucleus must inescapably include these ingredients, which come directly from the properties of the QCD Lagrangian.

There is little doubt that, at sufficiently
high nuclear density and/or temperature, quarks and gluons are the
correct degrees of freedom. By contrast, the general success of
conventional nuclear physics (with {\em effective} interactions)
indicates that nucleons and mesons provide a good starting point for
describing a nucleus at low energy. Therefore, a consistent nuclear
theory describing the transition from nucleon and meson degrees of
freedom to quarks and gluons is truly required to describe nuclei and
nuclear matter over a wide range of density and temperature.  Of
course, theoretically, lattice QCD simulations may eventually give
reliable information on the density and temperature dependence of
hadron properties in the nuclear medium.  However, current simulations
have just started to study the 2-body, nucleon-nucleon~\cite{LATN} and 
nucleon-hyperon interactions~\cite{LATY}, which is still very far from what is needed for
the description of a finite nucleus. At present, although
one is forced to rely on some models, it is a very important step to understand
the main features of nuclear phenomena and the structure of finite nuclei based on the quark and
gluon degrees of freedom~\cite{PPNP}.

We know that explicit quark degrees of freedom are certainly necessary
to understand deep-inelastic scattering (DIS) at momentum transfers of
several GeV. In particular, the nuclear EMC effect~\cite{EMC} has
suggested that it is vital to include some dynamics
beyond the conventional nucleon-meson treatment of nuclear physics to
explain the nuclear structure function data \cite{MILL-EMC,SMITH}.
This leads to the extraction of nuclear parton, or bound-nucleon parton
distributions~\cite{Eskola,Kumano}. In-medium structure functions were
studied at Jefferson Lab in an EMC-type experiment in heavy nuclei, emphasizing
the large-$x$ region~\cite{PR89008,PR02019}, and recently also in few-body
nuclei \cite{PR03103, Seely09}. Another series of Jefferson Lab experiments focus on the
longitudinal-transverse separation of in-medium structure functions 
\cite{PR99118,PR04001}.

Furthermore, the search for evidence of some
modification of nucleon properties in medium has recently been
extended to the nucleon electromagnetic form factors in polarized
(${\vec e},e^\prime{\vec p}$) scattering experiments on $^{16}$O
\cite{PR89033,Malov00} and $^4$He
\cite{Dieterich01,PR93049,PR03104,Strauch03, Paolone10} nuclei at MAMI
and Jefferson Lab.
These experiments observed the double ratio of proton-recoil
polarization transfer coefficients in the scattering off nuclei
with respect to the elastic $^4$He(${\vec e},e^\prime{\vec p}$)
reaction at four-momentum transfers squared of several 100~(MeV$/c)^2$
to a few (GeV/$c)^2$.
The results from $^4$He strongly hint at the need to include medium
modifications of the proton electromagnetic form factors.  The recent
Coulomb sum rule experiment at Jefferson Lab \cite{PR05110} is a precision
measurement of response functions in quasi-elastic electron scattering
and will provide important data on the nucleon charge properties in
the nuclear medium, providing important constraints on the
electric form factor of the proton.

The photoproduction of vector mesons on various nuclei has been
studied to search for possible in-medium modifications of the
vector-meson spectral functions.
The vector mesons, $\rho$, $\omega$, and $\phi$, are observed
{\it via} their decay to $e^+e^-$, in order to reduce the effects
of final state interactions in the nucleus
\cite{PR01112, Nasseripour07, Wood08}.
However, in these experiments mesons measured were moving relatively fast.
In-medium modifications at lower momenta have not yet been
observed experimentally. One can expect that the quarks inside
the fast moving mesons do not have enough time to adjust to the surrounding
nuclear medium. Thus, further experiments with nearly stopped meson kinematics
are certainly required to draw more concrete conclusions.

Another interesting class of medium modifications consists of the
changes in fundamental partonic processes that are embedded in the
medium. The well-established features of quark fragmentation in vacuum are
strongly modified in the cold nuclear medium, a topic best addressed
in semi-inclusive DIS on nuclei. A second effect is the alteration of
the process whereby a highly virtual quark radiates gluons, leading to
an increased rate of energy loss. Precise characterization of these effects in
the cold medium, in addition to giving new insight into QCD, may also
aid interpretation of observables in the hot medium. These
medium modifications of fundamental partonic vacuum processes are now
beginning to be understood theoretically after more than two decades
of investigation.

In this article, the medium modifications of nucleon structure, electromagnetic form factors
and structure functions are focused on in Section~2, while the modifications of meson structure,
masses and widths are focused on in Section~3. Section~4 treats the
medium modification of partonic processes, discussing modification of
fragmentation functions and medium-induced quark energy loss.
An overview of experiments at Jefferson Lab, dedicated to the study of medium
modifications of both nucleon and meson properties, is given in
Table~\ref{tab:experiments}.


\fulltable{\label{tab:experiments}Jefferson Lab experiments to
study medium modifications.}
\begin{tabular}{llp{4cm}l}
\br
In-medium property & Reaction & Target & Exp.~proposals \\
\mr
Form factors & $(\vec e, e'\vec p)$
& $^1$H, $^2$H, $^4$He, $^{16}$O
& \cite{PR89028, PR93049, PR03104, PR89033} \\
Nucleon charge & $(e,e')$
& $^4$He, $^{12}$C, $^{56}$Fe, $^{208}$Pb
& \cite{PR05110}\\
Structure function $F_2$ & $(e,e')$
& $^1$H, $^2$H, $^3$He, $^4$He, Be, C, Al, Fe, Cu, Au
& \cite{PR89008,PR02019,PR03103}\\
Structure functions $F_2$, $\sigma_L/\sigma_T$ & $(e,e')$
& $^1$H, $^2$H, C, Al, Fe
& \cite{PR99118,PR04001}\\
Vector-meson spectral functions & $(\gamma,e^+e^-)$
& $^2$H, $^{12}$C, Fe-Ti
& \cite{PR01112}\\
Fragmentation functions, $\Delta p_T$  & $(e,e'X)$
& $^2$H, C, Al, Fe, $^{120}$Sn, Pb
& \cite{PR02104} \\
\br
\end{tabular}
\endfulltable

\section{Nucleon structure modification}

\subsection{Modification of electromagnetic form factors}

\subsubsection{In-medium form factors}

Whether the nucleon changes its fundamental properties while embedded
in a nuclear medium has been a long-standing question in nuclear
physics, attracting experimental and theoretical attention.  
QCD is established as the theory of the strong
nuclear force but the degrees of freedom observed in nature, hadrons
and nuclei, are different from those appearing in the QCD Lagrangian,
quarks and gluons. There are no calculations available for nuclei
within the QCD framework. Nuclei are effectively and well described as
clusters of protons and neutrons held together by a strong, long-range
force mediated by meson exchange \cite{mosz60}.

Distinguishing possible changes in the structure of nucleons embedded
in a nucleus from more conventional many-body effects like
meson exchange currents (MEC), isobar configurations (IC) or
final state interactions (FSI) is only possible within the context
of a model.
Therefore, interpretation of an experimental signature as an
indication of nuclear medium modifications of the form factors is
better motivated if this results in a more economical description
of the nuclear many-body system.
In this context, a calculation by Lu {\it et al.}~\cite{Lu98}, using a
quark-meson coupling (QMC) model, suggests a measurable deviation from
the free-space electromagnetic form factor over the four-momentum
transfer squared $Q^2$ range $0 < Q^2 < 2.5$~(GeV/$c$)$^2$.
Similar measurable effects have been calculated in a light-front
constituent quark model by Frank {\it et al.} \cite{Fr96},
a modified Skyrme model by Yakshiev {\it et al.}~\cite{Ya02},
a chiral quark soliton model (CQS) by Smith and Miller \cite{Smith04},
and the Nambu-Jona-Lasinio model of Horikawa and Bentz \cite{Horikawa05}.
Medium modifications of nucleon properties in nuclear matter and
finite nuclei have been also discussed by Wen and Shen \cite{Wen08}.
Furthermore, the nuclear modification of axial form factors
\cite{qmcaxial} also may be measured.

The connection between the modifications induced by the nuclear medium 
of the nucleon form factors and of the deep inelastic structure 
functions was discussed by Liuti \cite{Liuti06} using the concept
of generalized parton distributions (GPDs).
Guzey {\it et al.} \cite{guzev} have studied incoherent deeply
virtual Compton scattering (DVCS) on $^4$He in the
$^4{\rm He}(e,e^{\prime}\gamma p)X$ reaction, which probes medium
modifications of the bound nucleon GPDs and elastic form factors.
They have also investigated medium modification of
the quark contribution to the spin sum rule~\cite{spinsum}.
The relation between the medium modified form factors and structure
functions was also discussed in Ref.~\cite{duality} in the framework of
quark-hadron duality, where the size of the medium modification in the
former was used to place constraints on models of the nuclear EMC effect
which assume a large deformation of the intrinsic structure of the
nucleon in medium.

The best experimental constraints on the changes in the electromagnetic form factors
come from the analysis of $y$-scaling data. For example, in the iron nucleus (Fe) the
nucleon root-mean-square radius cannot vary by more than 3\%~\cite{sick}.
However, in the kinematic range covered by this $y$-scaling analysis,
the $eN$ cross section is predominantly magnetic,
so this limit applies essentially to $G_{M}$.
(As the electric and magnetic form factors contribute typically in the
ratio 1:3 the corresponding limit on $G_{E}$ would be
nearer 10\%.) For the QMC model, the calculated increase
in the root-mean-square radius of the magnetic form factors is less than
0.8\% at $\rho_0$~\cite{Lu98}.
For the electric form factors the best experimental limit seems to
come from the Coulomb sum-rule, where a variation               
bigger than 4\% would be excluded~\cite{JOU}.
This is similar in size to the variations calculated in the QMC model
({\it e.g.}, 5.5\% for $G^p_E$ at $\rho_0$)~\cite{Lu98}
and not sufficient to reject them.

\subsubsection{Medium modifications from recoil polarization experiments}

One of the most intuitive methods to investigate the properties of
nucleons inside nuclei is quasi-elastic scattering off nuclei. Since
the charge and magnetic responses of a single nucleon are quite well
studied from elastic scattering experiments, measuring the same
response from quasi-elastic scattering off nuclei and comparing with a
single nucleon is likely to shed light on the question.
The polarization transfer ratio in elastic $\vec e p$ scattering,
$P'_x/P'_z$, is directly proportional to the ratio of the electric and
magnetic form factors of the proton \cite{Arnold81};
here $P'_x$ and $P'_z$ are the polarization transfer observables 
transverse and longitudinal to the momentum transfer direction.
When such measurements are performed on a nuclear target in quasi-elastic
kinematics, the polarization transfer observables are sensitive to the
form factor ratio of the proton embedded in the nuclear medium.
Experimental results for the polarization transfer ratio are
conveniently expressed in terms of the polarization double ratio
\begin{equation}
R = \frac{(P_x'/P_z')_{A}}{(P_x'/P_z')_{^1\rm H}},
\label{eq:rexp}
\end{equation}
in order to emphasize differences between the in-medium and free values.
Here the polarization transfer ratio for the quasi-elastic
proton knockout $A(\vec e, e'\vec p)$ reaction is normalized to the
hydrogen polarization transfer ratio measured in the identical
setting. Such a double ratio cancels nearly all experimental
systematic uncertainties.

\begin{figure}[h!]
\begin{center}
\includegraphics[width=\figwidth]{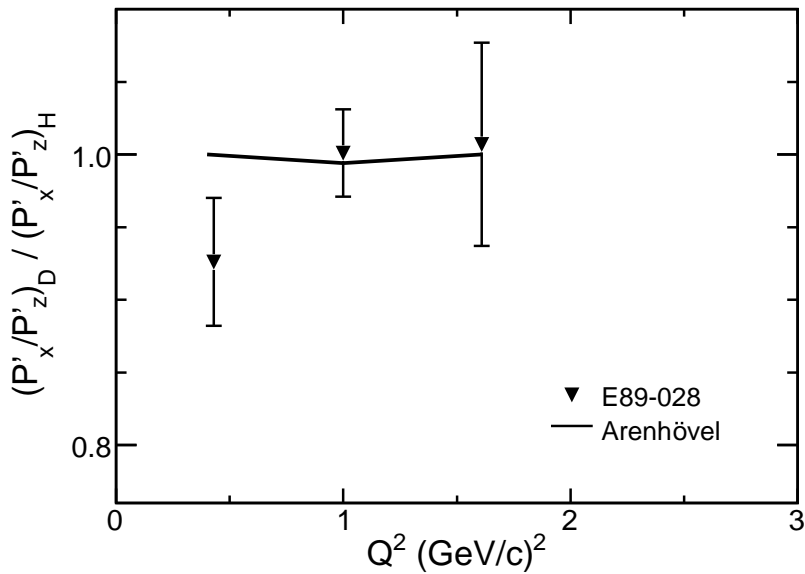}
\caption{
Bound-to-free polarization transfer double ratio $R$
for $^2$H$(\vec e, e' \vec p)n$ (triangles) from \protect\cite{Hu06}
as a function of $Q^2$. The curve shows results of a calculation
by Arenh\"ovel; see \cite{Hu06}.
\label{fig:ratio_2H}}
\end{center}
\end{figure}

Experiment E89-033 was the first to measure the polarization transfer in a
complex nucleus, $^{16}$O \cite{Malov00}.  The results are
consistent with predictions of relativistic calculations based
on the free proton from factor with an experimental uncertainty of
about 18\%. Earlier polarization transfer
experiments have studied nuclear medium
effects in deuterium \cite{Eyl95,Mil98,Bark99} at the Mainz microtron
(MAMI) and MIT-Bates. More recently, polarization transfer data on
$^2$H were measured in Jefferson Lab experiment E89-028 \cite{Hu06}. The data
are shown in Fig.~\ref{fig:ratio_2H} and
compared with a model calculation by Arenh\"ovel, 
which includes final state interactions,
meson exchange currents, and isobar configurations, as well
as relativistic contributions (RC) of leading order in $p/m_N$ to the
kinematic wave function boost and to the nucleon
current. Arenh\"ovel's calculation describes the $^2$H data well.  Within statistical uncertainties, no
evidence of medium modifications is found.  As
the sampled nuclear density is small and the bound proton is nearly on-shell
in the kinematics of this experiment, it is not surprising that there are no
indications for medium modifications of the proton electromagnetic
form factors in the $^2$H data.

One might expect to find larger medium effects in $^4$He, with its
significantly higher density.  Although estimates of the many-body
effects in $^4$He may be more difficult than in $^2$H, calculations
for $^4$He indicate they are small \cite{Laget94}. The first
$^4$He$(\vec e,e^\prime \vec p)^3$H polarization transfer measurements
were performed at MAMI at $Q^2 = 0.4$ (GeV/$c$)$^2$ \cite{Dieterich01}
and at Jefferson Lab Hall A at $Q^2$ = 0.5, 1.0, 1.6, and 2.6
(GeV/$c$)$^2$, E93-049 \cite{Strauch03}. Experiment E03-104
extended these measurements with two high-precision
points at $Q^2$ = 0.8 and 1.3 (GeV/$c$)$^2$ \cite{Paolone10}.  All these data were
taken in quasi-elastic kinematics at low missing momentum with
symmetry about the three-momentum transfer direction to minimize
conventional many-body effects in the reaction. The
missing-mass technique was used to identify $^3$H in the final
state.

\begin{figure}[tb!]
  \begin{center}
  \includegraphics[width=\figwidth]{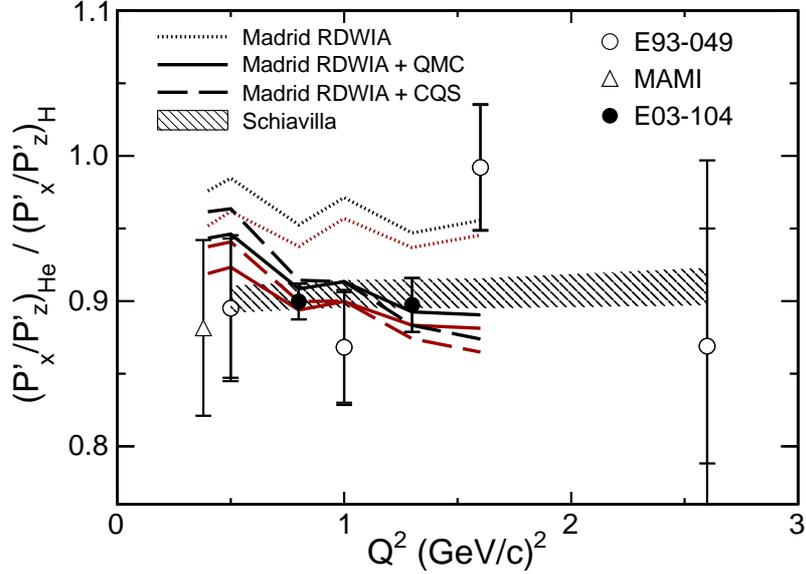}
  \caption{$^4$He$(\vec e, e'\vec p)^3$H
    polarization transfer double ratio $R$ as a function of
    $Q^2$ from Mainz \cite{Dieterich01} and Jefferson Lab experiments
    E93-049 \cite{Strauch03} (open symbols) 
    and E03-104 \cite{Paolone10} (filled circles).  The data are compared
    to calculations from Schiavilla {\it et al.}  \cite{Schiavilla05}
    and the Madrid group \cite{Ud98,Udias00} using the {\it cc1}
    (lower set of curves) and {\it cc2} (upper set of curves) current
    operators.  In-medium form factors from the QMC \cite{Lu98} (solid
    curves) and CQS \cite{Smith04} (dashed curves) models were used in
    two of the Madrid calculations.
    Not shown are a relativistic Glauber model calculation by the
    Ghent group \cite{Lava05} and results from Laget \cite{Laget94} which give
   both a value of $R\approx 1$.
    \label{fig:ratio}}
\end{center}
\end{figure}

The $^4$He polarization transfer double ratio is shown in Figure
\ref{fig:ratio}. The new data from E03-104 \cite{Paolone10}
are consistent with the previous data from E93-049 \cite{Strauch03}
and MAMI \cite{Dieterich01}. The polarization transfer ratio
$(P'_x/P'_z)$ in the $(\vec e,e' \vec p)$ reaction on helium is
significantly different from those on deuterium or hydrogen.
The helium data are compared with results of a relativistic
distorted-wave impulse approximation (RDWIA) calculation by the Madrid
group \cite{Ud98,Udias00} (dotted curve). In this model FSI are incorporated
using relativistic optical potentials that distort the final nucleon
wave function. MEC are not explicitly included in the Madrid
calculation. Predictions by Meucci {\it et al.} \cite{meucci} show
that the two-body current (the seagull diagram) effects are generally
small (less than 3\% close to zero missing momenta) and visible only
at high missing momenta.  It can be seen that the Madrid RDWIA
calculation (dotted curve) overestimates the data by about 6\%.
The calculations shown uses the Coulomb gauge, the {\it cc1} and
{\it cc2} current operators as defined in \cite{Forest83},
and the MRW optical potential of \cite{McNeil83}.
The {\it cc1} current operator gives lower values of
$R$ than the {\it cc2} operator.
In general, various choices for, {\it e.g.}, spinor distortions,
current operators, and relativistic corrections affect the theoretical
predictions by $\le$~3\% within the RDWIA model, and presently cannot
explain the disagreement between the data and the RDWIA calculation.
We note that these relativistic calculations provide good descriptions
of, {\it e.g.}, the induced polarizations as measured at Bates in the
$^{12}$C(e,e$^\prime \vec p$) reaction \cite{Woo98, Udias00} and of
$A_{TL}$ in $^{16}$O($e, e^\prime p$) as previously measured at
Jefferson Lab~\cite{Gao00}.
After including the density-dependent medium-modified form factors
from the QMC \cite{Lu98} or CQS \cite{Smith04} models in the RDWIA
calculation (solid and dashed curves), good agreement with the
polarization transfer data is obtained.

This agreement has been interpreted as possible evidence of proton
medium modifications \cite{Strauch03}. This interpretation is based on
the description of the data in a particular model in terms of medium
modifications of nucleon form factors and requires excellent control
of the reaction mechanisms such as meson exchange currents, isobar
configurations, and final state interactions.  In fact, there is an
alternative interpretation of the observed suppression of the
polarization transfer ratio within a more traditional calculation by
Schiavilla {\it et al.} \cite{Schiavilla05} (shaded band).
The latter calculation uses free nucleon form factors and explicitly
includes MEC effects which suppress $R$ by almost 4\%.
The FSIs are treated within the optical potentials framework and
include both a spin-dependent and spin-independent charge-exchange
term; the spin-orbit terms, however, are not well constrained by data.
In the model of Schiavilla {\it et al.}, the final state interaction
effects suppress $R$ by an additional 6\% bringing this calculation
also in good agreement with the data within the statistical
uncertainties associated with the Monte Carlo technique in this
calculation.
It should be noted that charge-exchange terms are not taken into account
in the Madrid RDWIA calculation. The difference in the modeling of
final state interactions is the origin of the major part of the
difference between the results of the calculations by Udias {\it et al.}
\cite{Ud98,Udias00} and Schiavilla {\it et al.}~\cite{Schiavilla05}
for the polarization observables.

\begin{figure}
\begin{center}
  \includegraphics[width=\figwidth]{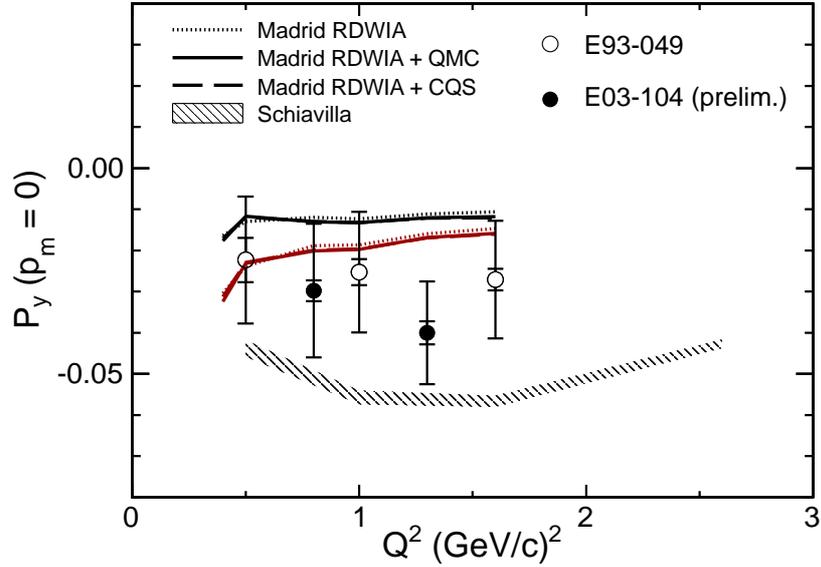}
  \caption{Induced polarization data from
    Jefferson Lab experiment E93-049 \cite{Strauch03} along with preliminary
    results from experiment E03-104 \cite{Malace08}. 
 The data are
    compared to calculations from the Madrid group \cite{Ud98,Udias00} and
    Schiavilla {\it et al.} \cite{Schiavilla05} as in Fig.~\ref{fig:ratio}.
Note that the comparison is made for missing momentum
    $p_m \approx 0$; the experimental data have been
    corrected for the spectrometer acceptance for this comparison.  }
  \label{fig:py}
\end{center}
\end{figure}

Effects from final state interactions can be studied experimentally with the
induced polarization, $P_y$, which is a direct measure of final state
interactions. Induced-polarization data were taken simultaneously to
the polarization transfer data.  Figure \ref{fig:py} shows the data
for $P_y$. The induced polarization is small at the low missing momenta in this reaction. The
sizable systematic uncertainties are due to possible instrumental
asymmetries.  Dedicated data have been taken during E03-104 to study
these and work is underway to significantly reduce the systematic
uncertainties in $P_y$ in the final analysis. The data are compared
with the results of the calculations from the Madrid group and
Schiavilla {\it et al.}  at missing momenta of about zero. To
facilitate this comparison, the data have been corrected for the
spectrometer acceptance.
The preliminary data suggest that the induced polarization (and thus
the final state interaction) is underestimated in the MRW optical
potential in the model of the Madrid group and overestimated in the
model of Schiavilla {\it et al.}
Note that the charge-exchange terms, particularly the
spin-dependent one, gives the largest contribution to P$_{y}$
in the Schiavilla {\it et al.} calculation. 
The induced polarization proves to be sensitive to the choice of
optical potential allowing this observable to be used to constrain
theoretical models of FSI.
A comparison of the model calculations in Figure \ref{fig:ratio} and
Figure \ref{fig:py} shows that the in-medium form factors mostly affect 
the ratio of polarization transfer observables and to a lesser extent
the induced polarization. It is a great advantage of E03-104 to have
access to both the polarization transfer and the induced polarization.

In summary, polarization transfer in the quasi-elastic $(e,e^\prime
p)$ reaction is sensitive to possible medium modifications of the
bound-nucleon form factor, while at the same time largely insensitive
to other reaction mechanisms. Currently, the $^4$He$(\vec e, e^\prime
\vec p)^3$H polarization transfer data can be well described by either
the inclusion of medium-modified form factors or strong
charge-exchange FSI in the models.
%
%
The analysis of the new high-precision induced-polarization data from
Jefferson Lab Hall~A has recently been finalized and the results provide 
now a more stringent test of these calculations \cite{Malace:2010ft}.

\subsubsection{Coulomb sum rule}

Experimental constraints on possible proton medium modifications are
available for the electric form factor $G_{E}$ from tests of the
Coulomb sum rule (CSR) in quasi-elastic electron scattering off nuclei.
This sum rule states that at large enough three-momentum transfer $q$
the integration of the charge response $R_L(q,\omega)$ of a nucleus over
the full range of energy loss $\omega$, excluding the elastic peak at
the lower limit, should be equal to the total charge $Z$ of the
nucleus \cite{Morgenstern01}
\begin{equation}
  S_L(q) = \frac{1}{Z} \int_{0^+}^{\infty}\frac{R_L(q, \omega)}{\tilde G^2_E} d\omega\;,
\end{equation}
where $\tilde G_E = (G^p_E + (N/Z) G^n_E)\zeta$ takes into account the
nucleon charge form factors and a relativistic correction, $\zeta$.
Pauli blocking at very small momentum transfer as well as short- and
long-range correlations at medium and larger momentum transfer result
in a quenching of the Coulomb sum rule.  However, at sufficiently
high momentum transfer, only short range correlation effects remain.
They have been estimated by various theoretical calculations using
different $NN$ forces and found to be responsible for at most 10\%
quenching of the CSR integral. As a result, any further quenching of
this quantity at sufficiently high momentum transfer may indicate the
possibility of modified properties of the nucleon
inside the nucleus~\cite{LResponse}.

Starting from this quite simple idea, for the past 20 years, various
laboratories around the world performed experiments in diverse conditions,
but the final conclusion is still controversial: on $^{4}$He, good
agreement between theory and experiment is obtained when using
free-nucleon form factors \cite{carlson} and similar results were
obtained on $^{12}$C and $^{56}$Fe \cite{JOU}.  However, a
re-examination of the extraction of the longitudinal and transverse
response functions on medium-weight and heavy nuclei ($^{40}$Ca,
$^{48}$Ca, $^{56}$Fe, $^{197}$Au, $^{208}$Pb and $^{238}$U) in the
framework of the Effective Momentum Approximation showed quenching
of the Coulomb sum rule  \cite{Morgenstern01,LResponse}.
This quenching corresponds to a
relative change of the proton charge radius of 13 $\pm$ 4\% and it
was interpreted as an indication for a change of the nucleon
properties inside the nuclear medium.

Jefferson Lab Experiment E05-110 is a precision test of the Coulomb
sum rule \cite{PR05110}. The experiment measured the inclusive
electron scattering cross sections off $^4$He, $^{12}$C, $^{56}$Fe,
and $^{208}$Pb in the quasi-elastic region and will make it possible
to study $A$- or density-dependent effects of the CSR. The measurement
covered a wide range of momentum transfers, 0.55~GeV/$c$ $\le q \le$
0.9~GeV/$c$, and thereby expanded the rather limited coverage of
previous experiments. This measurement will provide data on the
saturation or quenching of the Coulomb sum rule in a kinematic region
where a clean access of the nucleon charge properties in the nuclear
medium is plausible, free from Pauli blocking and long range
correlations. Also, in this region short-range correlations are found
to explain less than 10\% reduction from the naive Coulomb sum rule.
The kinematic range of the experiment overlaps with previous
experiments at the low-$q$ side and it provides a unique opportunity
to investigate the $q$ evolution of the Coulomb sum from the nucleonic
to sub-nucleonic scale.
Improved control of systematic uncertainties is possible through the
measurement of quasi-elastic scattering cross sections at four
different angles, providing four virtual photon polarization values
$\epsilon$ at almost equal spacing. Two additional $\epsilon$ settings
will provide a tool to study the effect of the Coulomb corrections.
Also, the use of two different thicknesses for $^{208}$Pb target will
help to reduce any systematic errors from the radiative corrections.
Data taking was completed in early 2008 and the analysis of the data,
particularly the extraction of the longitudinal and transverse
response functions, is underway.

\subsection{Modification of unpolarized structure functions}

The European Muon Collaboration (EMC) measured the ratios of structure
functions of iron and deuterium in deep-inelastic scattering and found
a reduction of the structure function of a nucleus compared to that of
a free nucleon at intermediate values of Bjorken $x$, $0.3 < x < 0.6$.
These measurements provided the first clear evidence that the quark
structure of nucleons and nuclei were significantly different.
Indeed, the nucleon bound in a nucleus carries less
momentum than in free space. However, the specific causes of the
modifications observed in nuclear structure functions have not yet
been identified with certainty \cite{Miller07}. Although the simple
kinematic effects of binding energy and Fermi motion do not account
for the EMC effect \cite{Miller07}, they do play a significant role at
large $x$ and are for those kinematics important in understanding the
modification of the nuclear quark distributions \cite{Arrington07}. A
detailed study of these effects, however, is hindered by limited data.
Most of the data are available only for heavy nuclei, yet few-body
nuclei have the advantage that exact calculations are available.
Particularly, uncertainties in the effect of binding due to
uncertainties in the nuclear structure are reduced. Calculations also
predict large differences in both the magnitude and $x$ dependence of
the EMC effect in $^3$He and $^4$He, whereas the effect in various
heavy nuclei show the same $x$ dependence. The study of few-body
nuclei can thus help to shed more light on the EMC effect. Data are
furthermore sparse above $x = 0.8$, where binding and other short-distance
effects should dominate. The $F_2$ structure function in inclusive
electron scattering off nuclei at large $Q^2$ has been measured at
Jefferson Lab Hall~C in a series of experiments:
E89-008 \cite{PR89008}, E02-019 \cite{PR02019}, and most recently
in E03-103 \cite{PR03103}.
The latter will provide better constraints on the effect of
binding in the large-$x$ region.

\begin{figure}[htb]
  \begin{center}
    \begin{minipage}[b]{0.48\linewidth}
\includegraphics[width=\fighwidth]{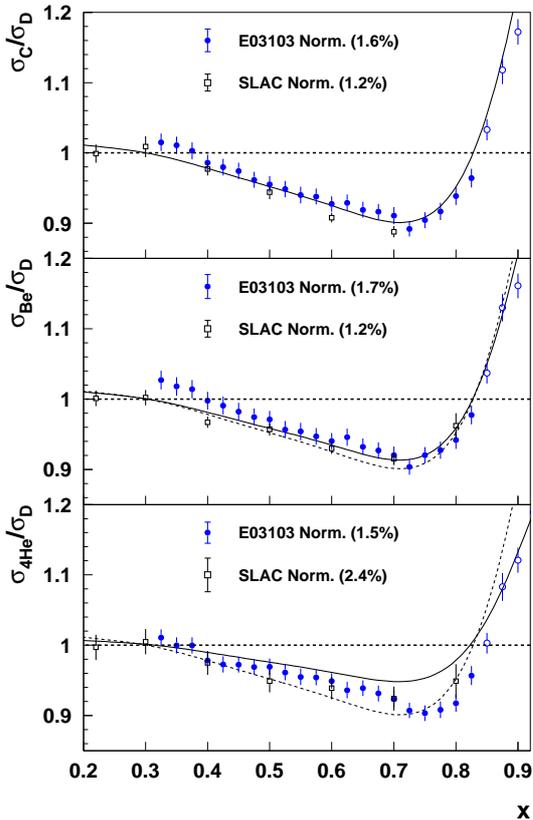}
    \end{minipage}\hfill
    \begin{minipage}[b]{0.48\linewidth}
      \noindent\caption{EMC cross-section ratios for $^{12}$C (top),
        $^9$Be, and $^4$He from E03-103 \cite{Seely09} and SLAC
        \cite{gomez94}. The $^9$Be results include a correction for
        the neutron excess (see \cite{Seely09}). Filled (open) circles
        denote $W^2$ above (below) 4~GeV$^2$. The solid curve is the
        $A$-dependent fit to the SLAC data, while the dashed curve is
        the fit to $^{12}$C. Normalization uncertainties are shown in
        parentheses for both measurements. Figure taken from
        \cite{Seely09}.\label{fig:emc}}
    \end{minipage}
  \end{center}
\end{figure}

Jefferson Lab Experiment E03-103 provides a precision measurement of the EMC
effect in both few-body nuclei and heavy nuclei. The experiment ran in
2004 in Hall C and measured inclusive electron scattering from $^1$H,
$^2$H, $^3$He, and $^4$He as well as Be, C, Al, Cu, and Au over a
broad range of $x$ ($0.3 < x < 1.0$) up to $Q^2 \approx 8.0$~GeV$^2$.
The $^3$He structure function was measured at $x$ larger than 0.5 for
the first time. Results from E03-103 along with previous SLAC results
are shown in Fig.~\ref{fig:emc} \cite{Seely09}. 
The Jefferson Lab data are in good agreement with these previous results and
improve significantly the precision of the EMC ratio at larger
$x$. The findings of E03-103 indicate that the nuclear dependence of
the cross section is comparable for $^4$He, $^9$Be, and $^{12}$C
(dashed curve); whereas the observed nuclear effects are clearly
smaller for $^3$He (not shown). The data thus do not support previous
$A$-dependent fits to the EMC effect nor fits dependent on the {\em
  average} nuclear density. The average nuclear density is relatively
low in $^9$Be. The results rather suggest that the nuclear dependence
of the quark distributions may depend on the {\em local} nuclear
environment. The better coverage at larger $x$ was achieved by
relaxing the usual $W^2$ cut for deep inelastic scattering from
4~GeV$^2$ down to below 2~GeV$^2$ after previous data showed that the
structure function scales also at lower values of $W$ (quark-hadron
duality) \cite{Arrington06}. Specific data were taken to verify this
assumption. Data from E02-019 will further allow a detailed study of
the scaling of nuclear structure functions in the $x>1$ region.

Several experiments in Hall C measured the longitudinal-transverse
separated structure functions $F_2$ and the ratio
$R=\sigma_L/\sigma_T$ from nuclear targets at $Q^2 < 4$~(GeV/$c$)$^2$
\cite{PR99118, PR04001}. The data on $R$ for hydrogen and deuterium
have been published \cite{Tvaskis07}, and the data on the heavier targets are
being analyzed. When fully available, these data will greatly increase
our understanding of nuclear effects.

\subsection{Modification of polarized structure functions}

The well-known EMC effect has fascinated a generation of physicists
and has stimulated much theoretical work. There is little question
that the explanation of this phenomenon involves the modification of
hadron structure in the medium, although the details of the
descriptions vary. It is then to be expected that there may be an
analogous change in the {\em{spin}} structure of the nucleon as
well. Detailed calculations support this conjecture \cite{pEMC} and
suggest that the observable modification is frequently larger than the
unpolarized EMC effect.

Letters of intent have proposed measurements at 6
GeV~\cite{LOI06003,LOI06004}, however, the best measurement will
be performed with the upgraded 12~GeV Jefferson Lab accelerator due to improved
kinematic reach and better control of systematic uncertainties. The
foundation that provides confidence that these studies can be realized
is the large and successful Jefferson Lab program of polarized target studies
on $^1$H, $^2$H, and $^3$He described elsewhere in this document.

The method requires polarizing a nuclear target that is
chosen to satisfy the following conditions:

\begin{itemize}
\item a large fraction of the overall nuclear polarization is
  concentrated on one nucleon within the nucleus;
\item reliable calculations can be performed of the degree of
  polarization for the highly polarized nucleon;
\item sufficient nuclear size to provide a medium environment
  representative of larger nuclei, while still limiting the dilution
  factor.
\end{itemize}

One example of a nucleus that satisfies these conditions
is $^7$Li, for which detailed nuclear structure calculations exist
\cite{rondon,ueda}. Asymmetry measurements can constrain the
value of the medium-modified $g_1$ to the few-percent level over the
range of interest in $x_B$ from 0.05 to more than 0.5.

\section{Meson structure modification}

The masses of hadrons are created dynamically and are much larger than
the summed masses of their constituents. The generation of hadronic
masses is connected to spontaneous breaking of chiral symmetry and
in-medium hadron properties, such as their masses and widths, are
expected to change with chiral symmetry restoration. Various
theoretical models predict an at least partial restoration of chiral
symmetry at high temperatures and/or densities.
By using effective chiral Lagrangians with a suitable incorporation of
the scaling property of QCD, Brown and Rho \cite{Brown91} derived
approximative in-medium scaling laws and predict for the ratio of
vector-meson masses at nuclear-matter density, $\rho_0$, compared to
their free values $m^*(\rho_0)/m \approx 0.80$. 
Using QCD sum rules, Hatsuda and Lee \cite{Hatsuda92} obtained the
density dependence of this ratio, $m^*(\rho)/m \approx 1 -
\alpha(\rho/\rho_0)$, with $\alpha = 0.18 \pm 0.06$. Models based on
nuclear many-body effects predict a broadening in the width of the
$\rho$ meson with increasing density. This prediction is based on the
assumption that many-body excitations may be present with the same
quantum numbers and can mix with the hadronic states
\cite{herrmann,rapp,mosel2,wambach,oset,madeline}.  Furthermore, due
to the uncertainty of the coupling constants as a function of density,
the branching ratios are expected to change in the nuclear medium and
also distort the invariant mass spectrum of the resonance
\cite{eichstaedt}.

First experimental studies of the properties of vector mesons as a
function of temperature and density were done in relativistic and
ultra-relativistic heavy-ion reactions. The CERES collaboration at CERN
performed pioneering investigations of possible $\rho$-meson medium
modifications through $e^+e^-$ pair production in nucleus-nucleus
collisions. A di-lepton yield exceeding expectations from hadron
decays has been observed in the mass region 0.2 to 0.6~GeV/c$^2$. An
upgraded CERES experiment with improved mass resolution confirmed
these results, and found in addition a substantial in-medium
broadening of the $\rho$ spectral function over a density dependent
shift of the $\rho$ pole mass \cite{Adamova08}. The result of
$\mu^+\mu^-$ measurements in In-In collisions in the NA60 experiment
at CERN has also shown a considerable broadening (doubling of the
width) of the $\rho$ spectrum, while no shift in the mass was observed
\cite{na60,na60-1,na60-2}. In these heavy-ion reactions the
temperature and/or density is varied and they proceed far from
equilibrium under difficult to separate reaction mechanisms. This makes an
interpretation of these results in terms
of the chiral symmetry structure of the vacuum very challenging.

Medium modifications of mesons at zero temperature and normal nuclear
matter density are experimentally accessible in photonuclear or
elementary hadronic reactions on heavy nuclei. The best approach, free
of final state interactions, is the study of the leptonic decay
channel of these mesons. An observation of a medium-modified vector
meson invariant mass decrease ($\alpha = 0.09 \pm 0.002$) has been
claimed by a KEK-PS collaboration in an experiment where 12~GeV
protons were incident on nuclear targets, and the $e^+e^-$ pairs were
detected~\cite{kek,kek2,kek-new}. The Crystal
Barrel/TAPS collaboration has reported a 14\% downward shift in the
mass of the $\omega$, where the analysis focused on the $\pi^0\gamma$
decay of {\it low-momentum} $\omega$ mesons photoproduced on a nuclear
target~\cite{taps}. 
However, in a recent re-analysis of these data the earlier claim of an
in-medium mass shift of the $\omega$ could not be confirmed \cite{Nanova10}. 
It is predicted that the $\omega$ potential in nuclear matter
is attractive and possible to form nuclear bound state~\cite{qmcomega,hayano,qhdomega},
if it is produced nearly with zero momentum.


The Jefferson Lab Experiment E01-112 studied the photoproduction of light
vector mesons ($\rho$, $\omega$, and $\phi$) on deuterium and the
heavier targets of carbon, titanium, and iron \cite{PR01112} with the
CEBAF Large Acceptance Spectrometer (CLAS) \cite{Mecking03}. The photon
beam from the Hall B photon tagging facility covered an energy range
from about 0.6 to 3.8~GeV. The CLAS has an excellent electron/pion
discrimination, which allowed detection of the vector mesons {\it via}
their rare leptonic decay in the $e^+e^-$ channel and thus avoiding
the complications from final state interactions of the hadronic decay
channels.

The experimental $e^+e^-$ invariant mass distribution includes the
spectral distribution from the decay of the vector mesons $\rho$,
$\omega$, and $\phi$. It also contains background from other physical
processes, for example meson decays into $\gamma e^+e^-$ or the
Bethe-Heitler process.
Detailed simulations showed that from all background
processes, which lead to correlated $e^+e^-$ pairs, only the Dalitz
decay of the $\omega$ mesons contributes appreciably in
the $e^+e^-$ mass region of interest.
To simulate each physics process, a realistic model was employed and
corrected for the CLAS acceptance. The events were generated using a
code based on a semi-classical Boltzmann-Uehling-Uhlenbeck (BUU)
transport model developed by the Giessen group that treats
photon-nucleus reactions as a two-step process~\cite{mue02}. In the
first step, the incoming photons react with a single nucleon, taking
into account various nuclear effects, {\it e.g.} shadowing, Fermi motion,
collisional broadening, Pauli blocking, and nuclear binding.  Then, in
the second step, the produced particles are propagated explicitly
through the nucleus allowing for final state interactions, governed by
the semi-classical BUU transport equations. A rather complete
treatment of the $e^+e^-$ pair production from $\gamma A$ reactions at
Jefferson Lab energies using this code can be found in
Ref.~\cite{bra99}.

Another source of background for the $e^+e^-$ mass distributions, the
combinatorial background, are events where the electron and the
positron originate in uncorrelated processes within the same 2 ns wide
CEBAF beam bucket. The most salient feature of the uncorrelated
sources is that they produce same-charge lepton pairs as well as
oppositely charged pairs. This is also true for the measurement of
opposite-sign pions or muons for which the combinatorial method has
also been used in the past \cite{pions,muons}.  This method has also
been used in the extraction of resonance signals \cite{res} and proton
femtoscopy of $eA$ interactions \cite{stepan}. The combinatorial
background is determined by an event-mixing technique. The electrons
of a given event are combined with positrons of another event, as the
two samples of electrons and positrons are completely uncorrelated.
This produces the phase-space distribution of the combinatorial
background.  This distribution was then normalized to the number of
expected opposite-charge pairs, which unambiguously can be determined
from same-charge pairs.

\begin{figure}[ht]
\begin{center}
\includegraphics[width=\fighwidth]{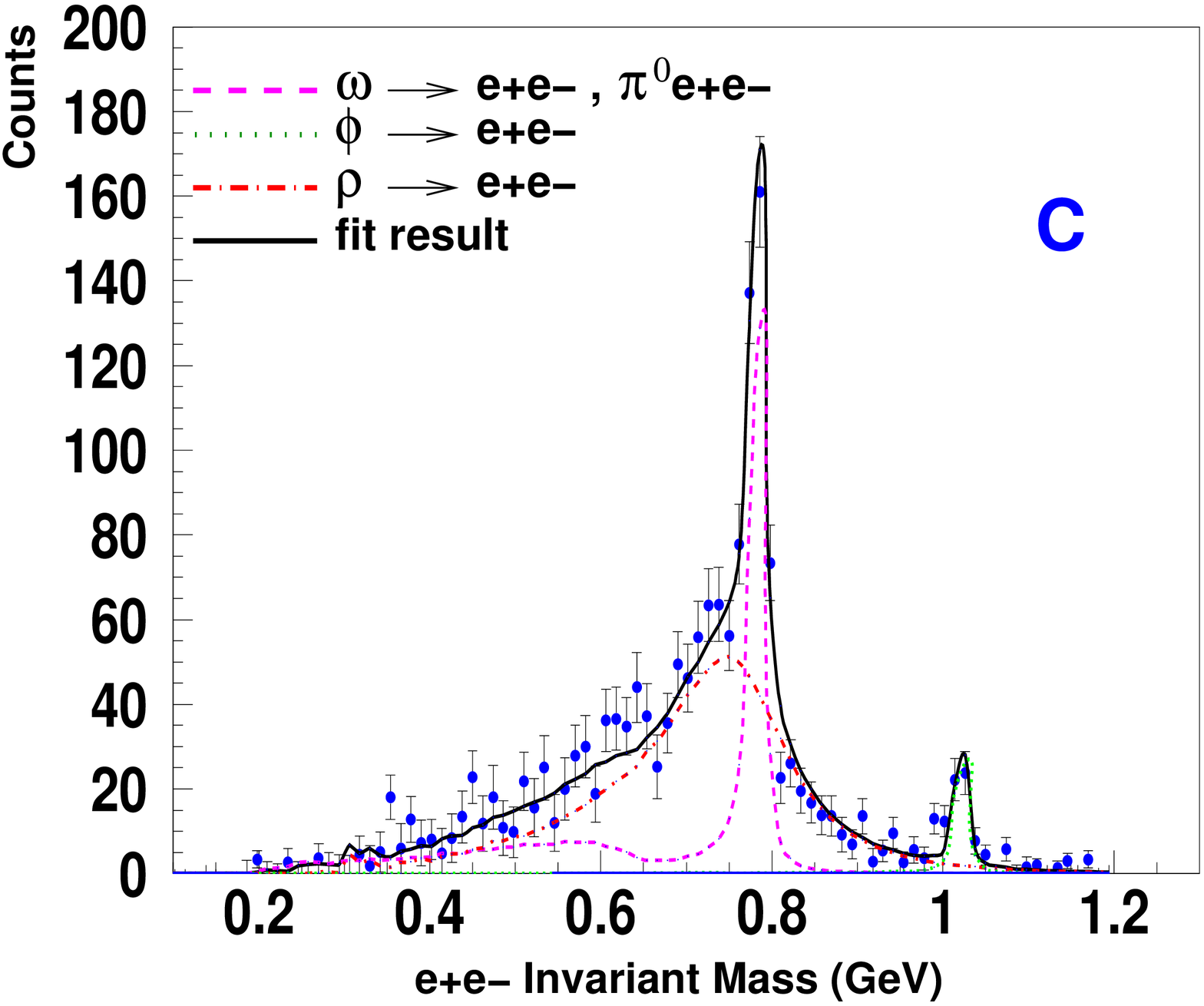}
\includegraphics[width=\fighwidth]{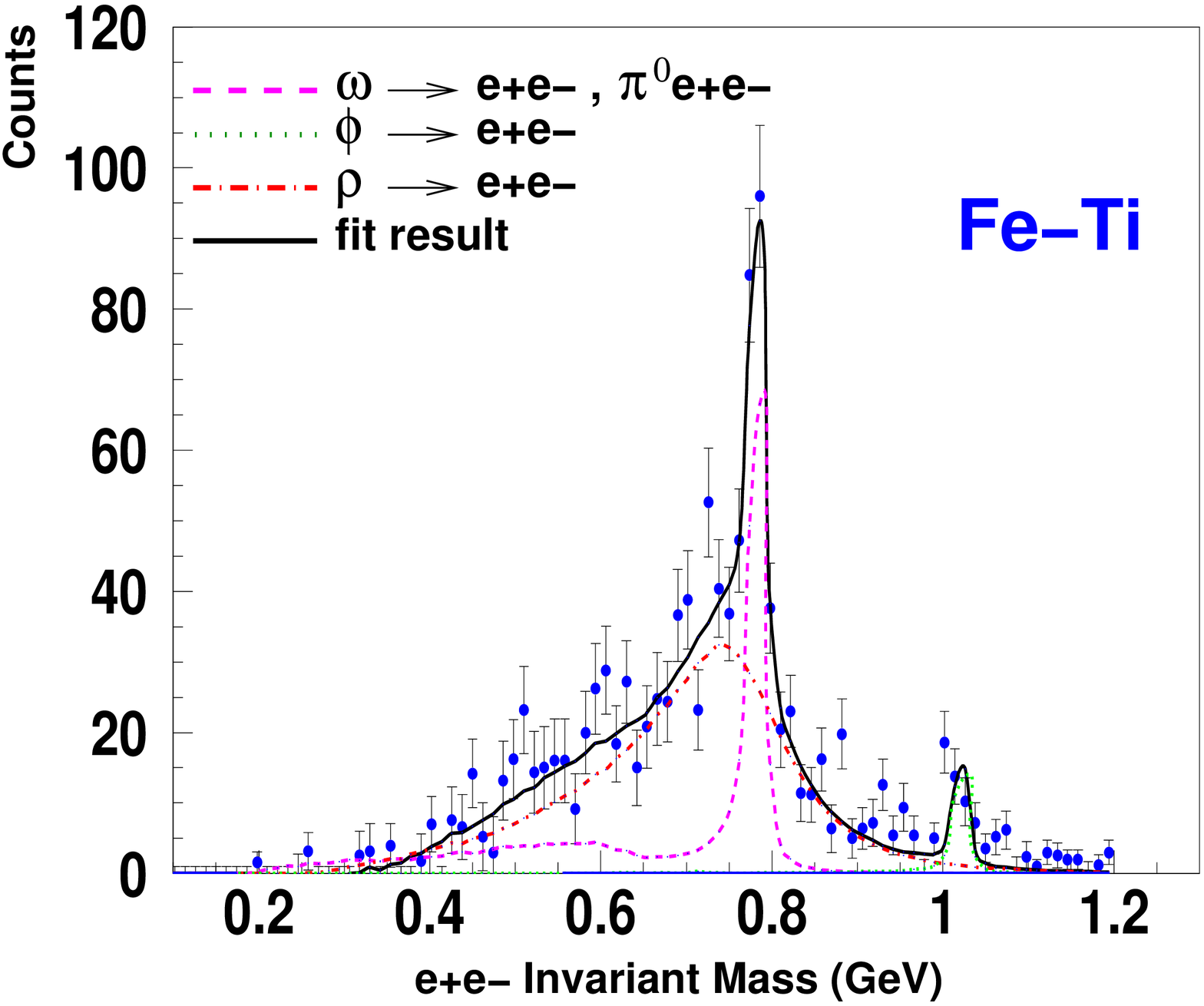}
\caption{Experimental $e^+e^-$ invariant mass distributions obtained
  for carbon (left panel) and Fe-Ti (right panel) after
  subtraction of the combinatorial background. The curves are
  calculations by the BUU model \cite{effenberger1,effenberger2} for
  various $e^+ e^-$ channels. Figures are from \cite{Nasseripour07}.}
\label{fig:g7}
\end{center}
\end{figure}

\begin{figure}[ht]
\begin{center}
\includegraphics[width=\fighwidth]{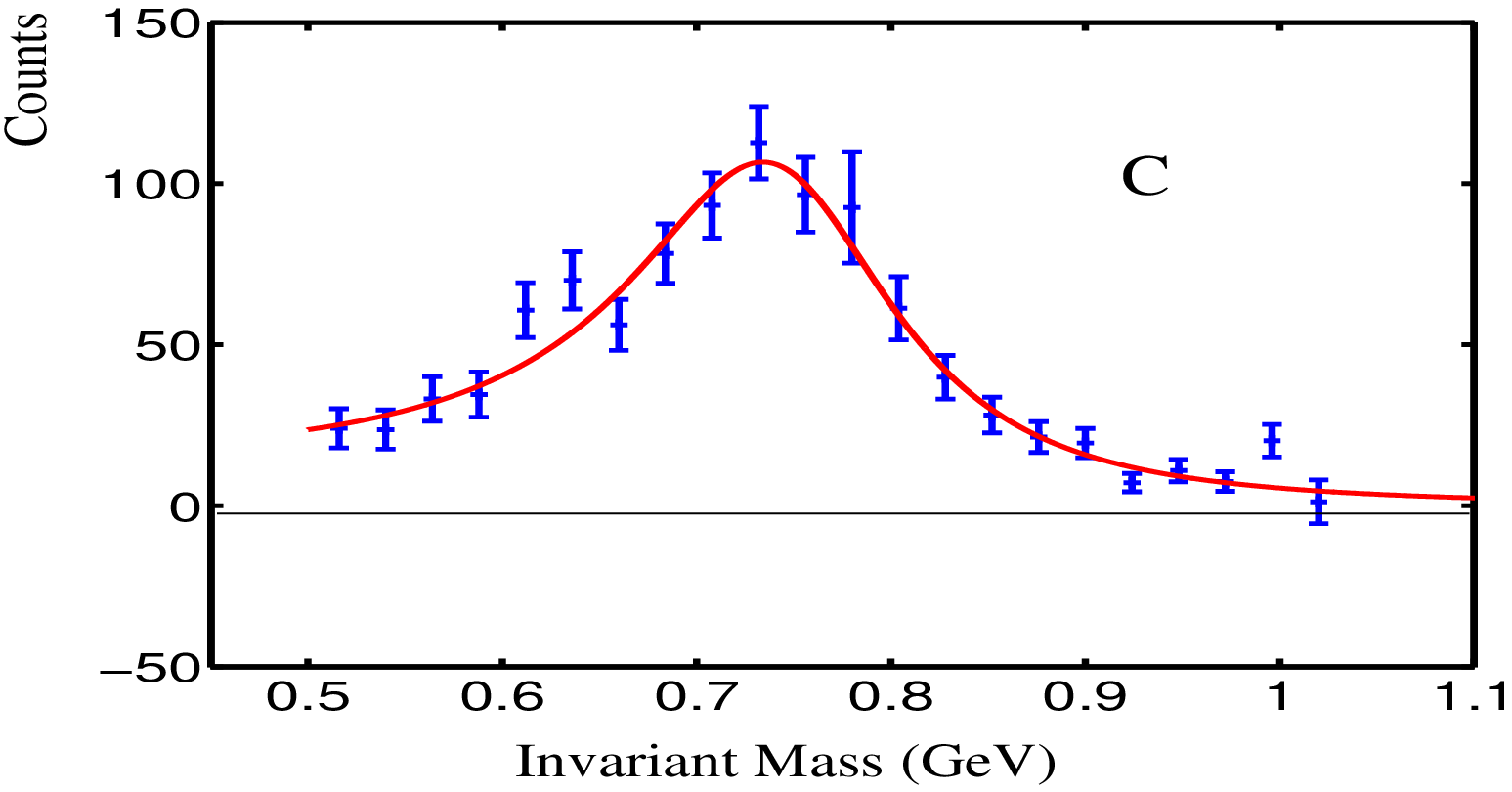}
\includegraphics[width=\fighwidth]{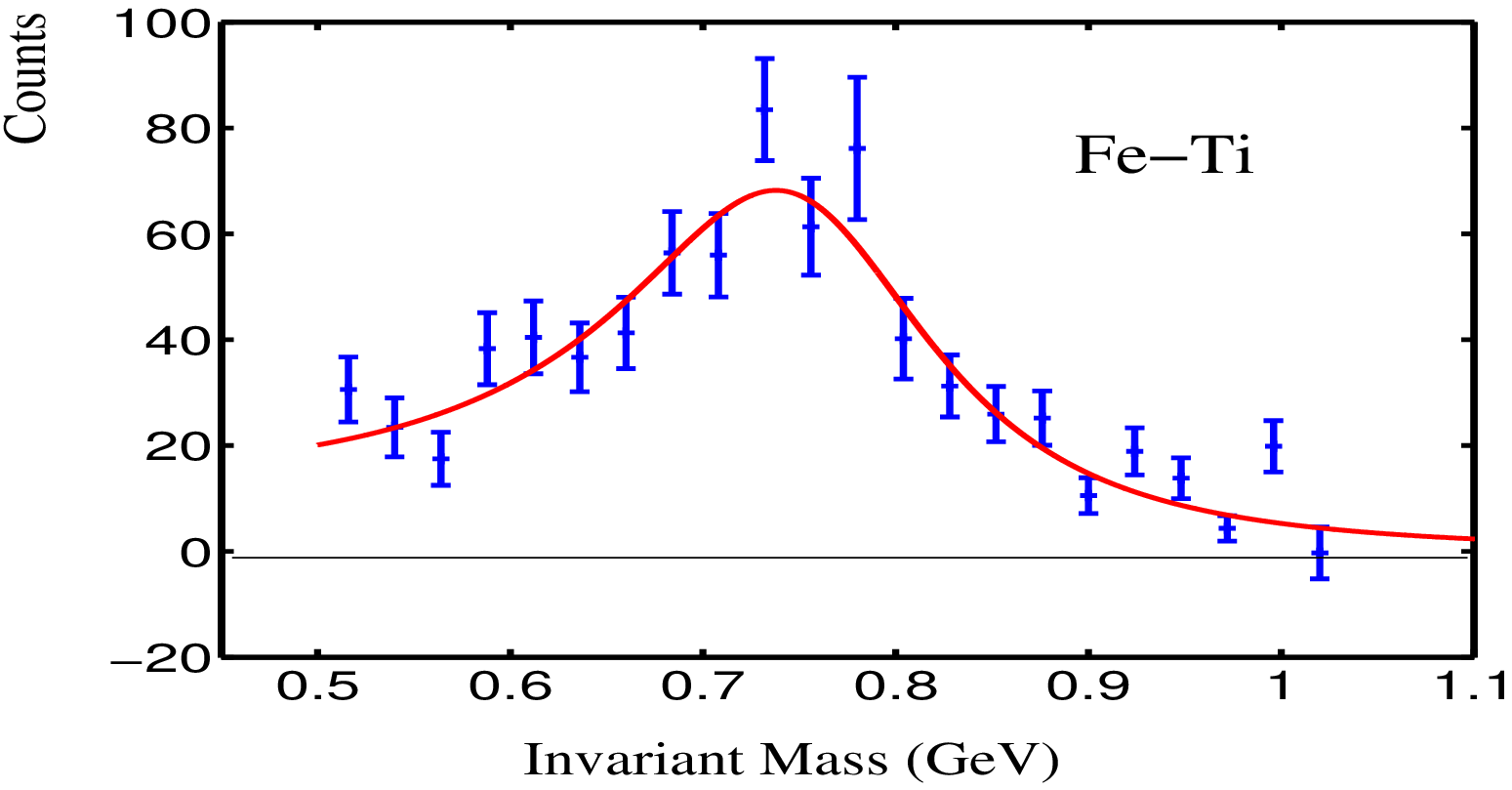}
\caption{Extracted $\rho$ mass spectra from the CLAS g7 run data
  for the C (left), and Fe-Ti (right panel) targets.
  The curves show fits of a
  Breit-Wigner function times $1/\mu^3$, where $\mu$ is the $e^+e^-$
  invariant mass. Figures are from \cite{Nasseripour07}.}
\label{fig:g7-fit}
\end{center}
\end{figure}

The experimental $e^+e^-$ mass distributions are shown in
Fig.~\ref{fig:g7} after subtraction of the combinatorial
background. Also shown are BUU model calculations for various $e^+e^-$
channels, fitted to the data, which describe the data very well. The
narrow distributions of the $\phi$ and $\omega$ meson, including its
Dalitz decay, are readily normalized to the data. The subtraction of
these distributions results in the experimental spectra of the $\rho$
mass shown in Fig.~\ref{fig:g7-fit}. The shape of the spectral
function is well approximated by a Breit-Wigner function times a
factor $1/\mu^3$, where $\mu$ is the mass of the $e^+e^-$ pair
\cite{Wood08}. Experimental values of the in-medium $\rho$ mass and
width are then obtained from fits to the data (solid curves).  The
in-medium widths of the $\rho$ for the nuclear targets are slightly
larger than the free value, {\it e.g.} $\Gamma = 218 \pm 15$ MeV for
the Fe-Ti target, but are well understood as collisional
broadening \cite{bugg} as modeled in the BUU calculations. They
are not compatible with the doubling of the $\rho$ width
reported by NA60 \cite{na60}.

The observed in-medium $\rho$ masses are consistent with the free
value. The relative $\rho$-mass shift for the Fe-Ti target and $\rho$
momenta ranging from 0.8 to 3.0~GeV is $\alpha = 0.02 \pm 0.02$. This
is consistent with the predictions of no significant mass shift by the
calculations of Ref.~\cite{wambach,oset} and those of
Ref.~\cite{mue02,mue04} at $\rho$ vector meson momenta $>$ 1~GeV. The
total systematic uncertainty for the measured $\alpha$ due to various
sources is estimated to be $\Delta \alpha = \pm 0.01$ \cite{Wood08}.
The result from the CLAS g7 run sets an upper limit of $\alpha = 0.04$
with a 95$\%$ confidence level. It does not favor the prediction of
Refs.~\cite{Brown91} and \cite{Hatsuda92} for a 20$\%$ mass shift and
$\alpha = 0.16 \pm 0.06$ respectively, and is significantly different
from other similar experiments~\cite{kek,kek2,kek-new}, where $\alpha
= 0.092 \pm 0.002$, with no broadening in the width of the $\rho$
meson. The g7 results are consistent with the
result of a recent re-analysis of $\omega$ meson photoproduction
data \cite{Nanova10} which found broadening, but no mass
shift, in the $\omega$ signal.

The extracted experimental $\rho$ mass spectrum with the unique
characteristic of electromagnetic interactions in both the production
and decay is well described by the $\rho$ functional form obtained
from the exact calculations given in Refs.~\cite{bra99,guo,oconnell}
with no modification beyond standard nuclear many-body effects.
Recently, the dilepton spectra were calculated by Riek {\it et al.}~\cite{Riek2009,Riek2010} within the same effective hadronic model that describes
 NA60 data \cite{vanHees:2007th} and found to be in good agreement with the CLAS data.
With
the availability of more sophisticated theoretical models and improved
analysis techniques, future experiments with higher statistics are
expected to make a conclusive statement about the momentum dependence
of the in-medium modifications and the nature of the QCD vacuum.

\section{Modification of partonic processes}
A tremendous amount of progress has been achieved over the past decade in
understanding the modification of partonic processes in the medium. New experiments
at Fermilab, DESY, and Jefferson Lab have uncovered a wealth of new information on partonic processes
in the {\em{cold}} nuclear medium through the Drell-Yan process and through semi-exclusive DIS. Experiments at RHIC
have generated a strong interest in the same partonic processes in the {\em{hot}} medium
through the observation of jet quenching, and these observations will soon be tested with much higher energies at the LHC~\cite{JetQuenching}.
The Jefferson Lab experiments~\cite{PR02104,PR1206117} in particular promise
to uniquely measure spacetime properties of QCD inaccessible to any other experiments, with sufficient
luminosity and kinematic reach to achieve a detailed understanding of the processes by which hadrons are formed
and of the propagation of quarks both in the medium and in vacuum.

A primary aim of the measurements in the cold medium is to explicate the fundamental QCD
processes involved in parton propagation and hadron formation. Particularly in the
space-time domain, the well-understood properties of nuclei can be exploited to extract quantities such as the
free quark lifetime (the so-called production time) and hadron formation times. In the process of confronting the
data with model calculations, the detailed mechanisms of hadron formation can also be constrained and characterized.
As a secondary benefit, the insight gained by these studies can in principle be used to better understand the data
from the relativistic heavy ion collisions. Much theoretical work is needed to actually accomplish this, however, there are examples of efforts to describe both the hot and cold matter within the same theoretical language \cite{Majumder08a,Majumder08b}.

The following discussion centers around two primary interests. First, the modification of fragmentation functions is discussed.
The primary observable, the hadronic multiplicity ratio, is a measure of the modification of the fragmentation
functions. Through modeling it can be related to hadron formation lengths and fragmentation mechanisms. Second, the topic
of quark energy loss is discussed. Here the primary observable is the broadening of the distributions in hadron
transverse momentum. It is expected that the production time can be estimated from this variable within certain kinematic
regions. Further, this observable can be connected to a variety of important and interrelated topics, such as
partonic multiple scattering in the medium, energy loss {\it via} gluon bremsstrahlung, and jet quenching in
relativistic heavy ion collisions. Recent reviews, with a more complete coverage of theoretical approaches, may be found in
Refs.~\cite{Accardi2009,Majumder10}. 

\subsection{Fragmentation functions}

In the simplest picture, fragmentation functions $D^q_h(z)$ describe
the probability that an energetic quark $q$ or antiquark $\bar{q}$
evolves into a specific hadron $h$, where the ratio of the hadron
energy to the quark energy is $z=E_h/E_q$. They enter into the
expression for the cross section for semi-inclusive electron
scattering in combination with the parton distribution functions. A
direct measure of the nuclear medium modification of the fragmentation
functions is given by the {\em{hadronic multiplicity ratio}} $R_M^h$:

\begin{eqnarray}
\label{eq:had_mult_rat}
R_M^h(z,\nu,Q^2,p_T^2,\phi)
=
\frac{\Biggl( \frac{N_h(z,\nu,Q^2,p_T^2,\phi)}{N_e^{DIS}(\nu)}\Biggr)_A}
{\Biggl( \frac{N_h(z,\nu,Q^2,p_T^2,\phi)}{N_e^{DIS}(\nu)}\Biggr)_D},
\end{eqnarray}

\begin{figure}[tb]
\begin{center}
\includegraphics[width=0.49\linewidth]{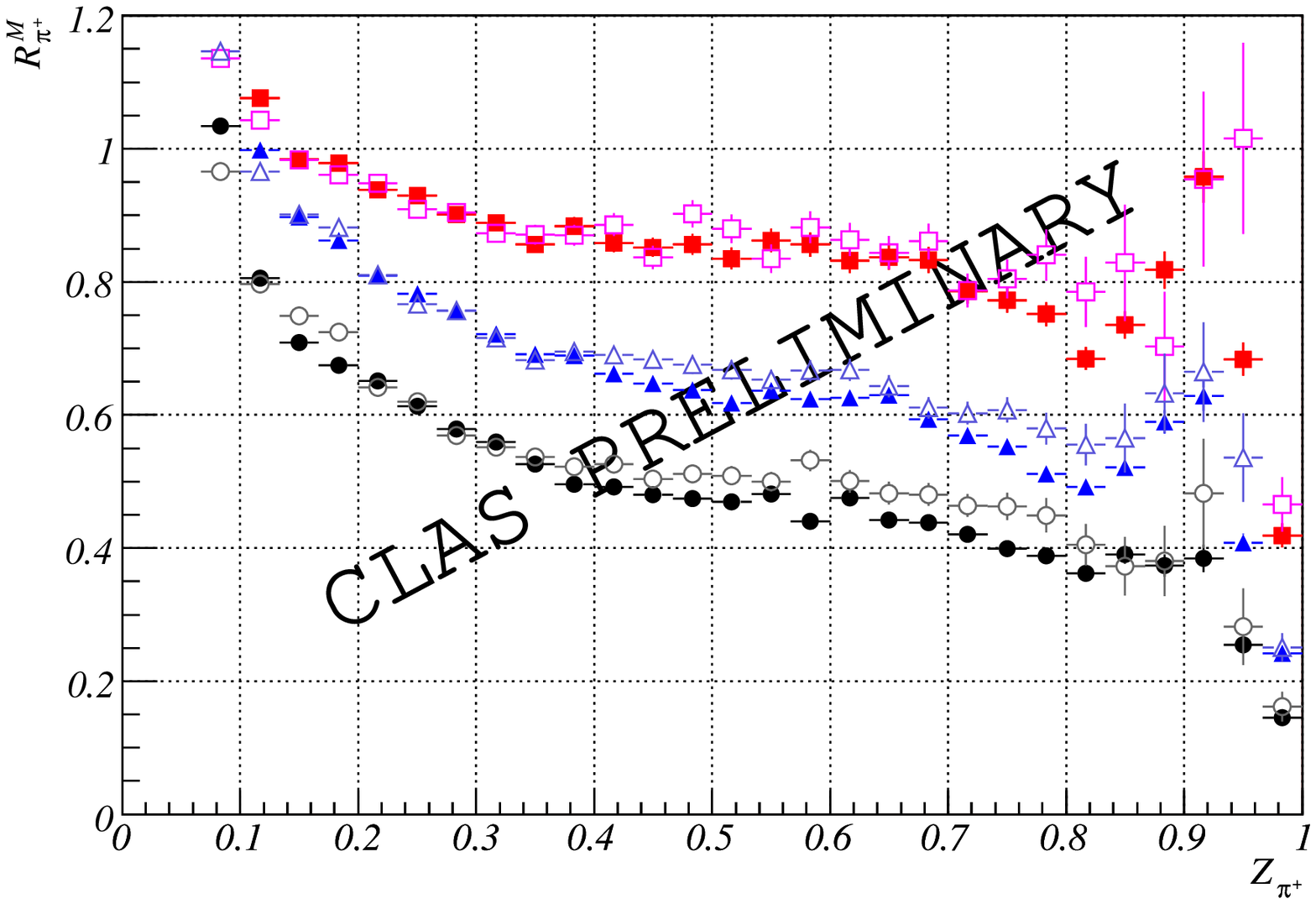}
\includegraphics[width=0.49\linewidth]{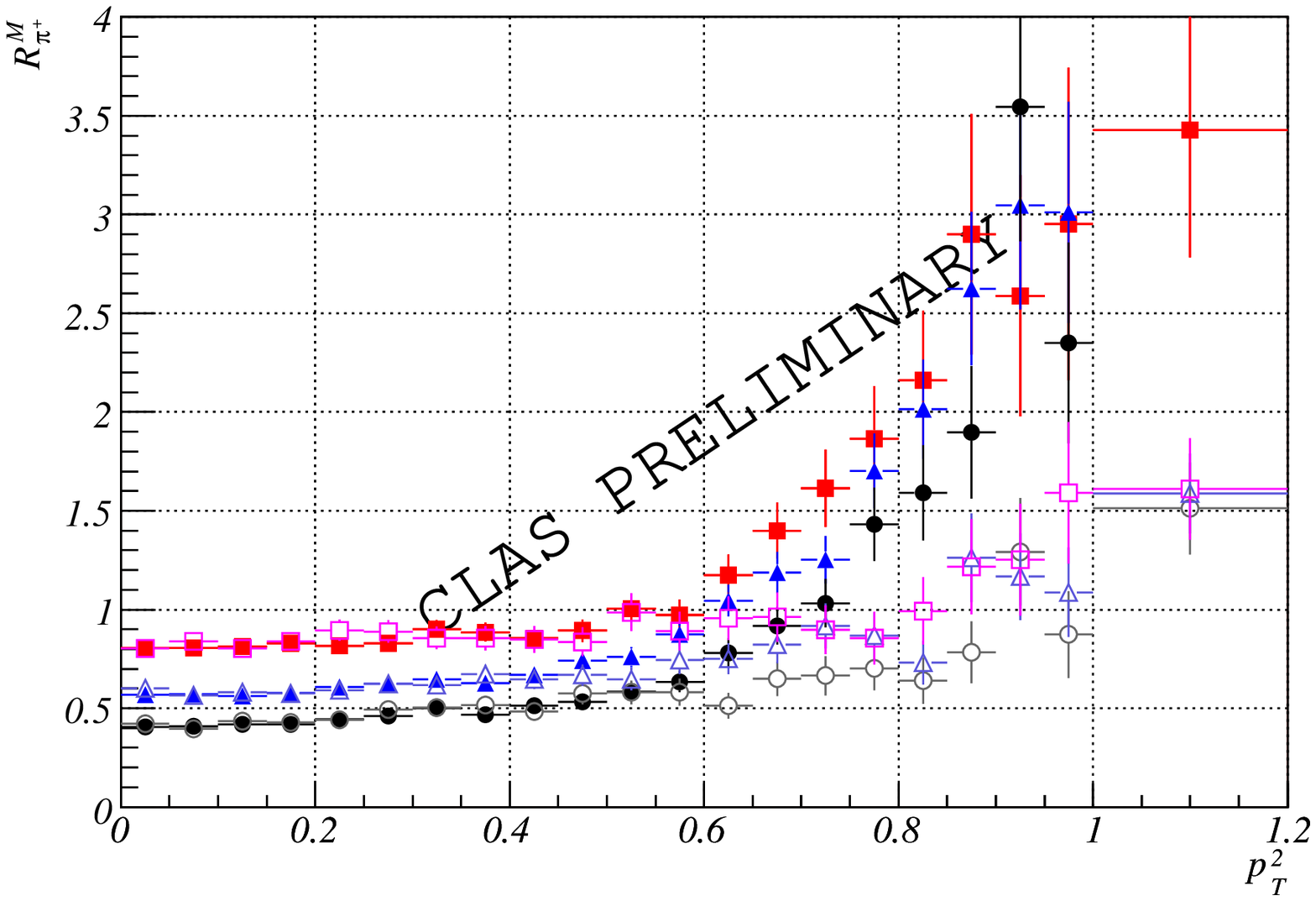}
\caption{Preliminary data for the CLAS hadronic multiplicity ratio
  measurement as a function of $z_{\pi^+}$ (upper plot) and $p_T^2$ (lower
  plot). The data shown are for positive pions for carbon (squares),
  iron (triangles), and lead (circles) targets normalized to a deuterium target. The data
  with filled (open) symbols correspond to the range
  $\nu=2.2-3.2$ ($3.2-3.7$)~GeV and $Q^2=1.0-1.3$~GeV$^2$. The errors
  shown are statistical only. See Refs.~\cite{Hakobyan08,Brooks08}
  for more discussion of the preliminary results for positive pions,
  and Ref.~\cite{Hicks09} for preliminary results for neutral kaons.
\label{fig:CLAS_R_z_pt2}}
\end{center}
\end{figure}

\noindent
where the numerator pertains to larger nucleus $A$, the denominator
pertains to the deuterium nucleus $D$, and the variables not
previously defined are $p_T^2$, the hadron momentum transverse to the
virtual photon direction squared, and $\phi$, the azimuthal angle
around the virtual photon direction of the electron-hadron scattering
plane. This observable, from the parton model perspective, is
determined by the parton distribution functions and the fragmentation
functions in the two nuclei. Experimentally it is now known that the
nuclear medium induces a strong $z$ dependence on $R^h_M$,
demonstrating that it is indeed the fragmentation functions that are
modified, in addition to the much smaller changes in the parton distribution
functions that are associated with the EMC effect.

While there were historical measurements of semi-inclusive hadron
production on nuclear targets prior to the 1990's, the first
measurements that included identification of the produced hadrons were
those of the HERMES experiment at DESY. These pioneering measurements
established the basic features of the multiplicity ratio:

\begin{itemize}
\item suppression of $R^h_M$ at high $z$ and at low $\nu$ that
  systematically increases for larger nuclei;
\item an enhancement of $R^h_M$ at high $p_T^2$ in analogy with the
  Cronin effect seen in $p-A$ collisions;
\item $R^h_M$ depends on the hadron species produced.
\end{itemize}

The Jefferson Lab 5~GeV experiment, while more limited in kinematic reach than
the HERMES experiment, has been able to observe qualitatively new
behavior due to the much higher Jefferson Lab luminosity and the capacity for
accommodating solid targets, thus probing the largest nuclei. Because
the Jefferson Lab data have two orders of magnitude more integrated luminosity, it
is possible to bin the data in up to three kinematic variables while
preserving good statistical accuracy. Thus, a fuller exploration of
the kinematic dependence of $R^h_M$ is possible. A second consequence
of the higher luminosity is access to production of rarer hadrons,
such as $K^0$, $\eta$, and potentially $\eta'$, as well as exploratory studies with baryons such as protons, $\Lambda$s and $\Sigma$s.
This complements the
HERMES data set, in which the RICH detector was able to identify $K^+$
and $K^-$ as well as antiprotons; both experiments have access to the
three pion charge states, which provides a good cross-check to
validate the consistency of the two data sets. Figure~\ref{fig:CLAS_R_z_pt2}
shows a small subset of the preliminary Jefferson Lab data for $R^h_M$ for positive pions. It can
be seen from this figure that the basic dependencies established by
the HERMES data are well reproduced by these lower energy data
(suppression at high $z$, enhancement at high $p_T^2$), and
are extended to the heaviest nuclei and to three-fold differential
distributions ($\nu$, $Q^2$, and $z$ in the upper panel, and $\nu$,
$Q^2$, and $p_T^2$ in the lower panel). In addition to these studies
for positive pions, preliminary results for neutral kaons can be found
in Ref.~\cite{Hicks09}, and studies are underway for neutral pions
and $\eta$~\cite{Mineeva09}, negative pions, protons, and positive
kaons~\cite{Dupre10}, and $\Lambda$ and $\Sigma$ baryons.
These types of studies lay the groundwork for future experiments at
high luminosity facilities \cite{PR1206117, BrooksEIC}.

To interpret these data and extract the physics of interest, a model
is needed. Many models have been constructed to describe the HERMES
results, and a number of these are now being used to address the new
Jefferson Lab data.
Existing models fall into three classes. First, there are models
which assume that hadron formation only takes place outside the nucleus,
and then use a pQCD framework to describe the interaction of a moving
parton with the medium. In this picture, the behavior of $R^h_M$ is
due entirely to the stimulated emission of gluons resulting from
partonic multiple scattering. The second class of models invokes
hadron formation within the medium as the primary cause of the behavior
of $R^h_M$, such that the semi-classical interaction of a forming
hadron (or `prehadron') removes flux from the observed final state
channel through hadronic interactions.
The third class of model has emerged only
recently; it retains ingredients from the other two approaches, but
treats the interaction in a fully quantum-mechanical framework. In
this last approach an interference is observed between hadron formation
inside the medium and outside the medium, and this gives rise to a
non-trivial modification of $R^h_M$.

It is likely that the stringent constraints imposed by the new Jefferson Lab
data will soon clarify what the essential model ingredients are.
In particular, it is crucial to establish the relative importance of
the two processes of gluon bremsstrahlung and prehadron interactions
in describing the kinematic and flavor dependencies of $R^h_M$, and to
understand the role of quantum interference in these processes.
At that point it will be feasible to fully analyze the data from the
5~GeV data and from the planned 12~GeV Jefferson Lab Experiment
E12-06-117, extracting hadron formation lengths from a variety of
produced hadrons and constraining the mechanisms involved in their
formation.
\begin{figure}[tb]
\begin{center}
\includegraphics[width=\textwidth]{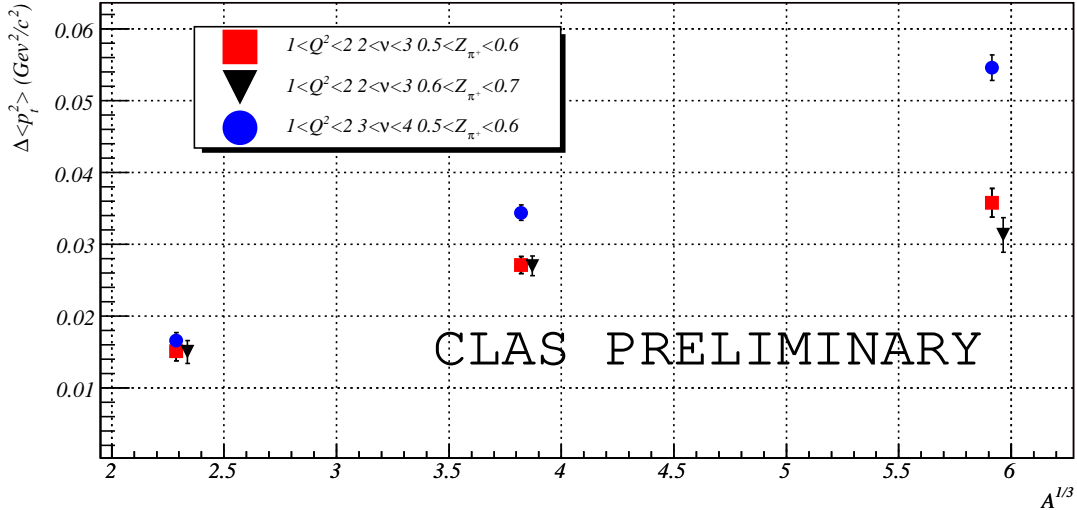}
\caption{Mass number dependence of $p_T$-broadening in nuclear DIS. CLAS preliminary data for positive
  pions for carbon, iron, and lead targets three-fold differential in
  $Q^2$, $\nu$, and $z_{\pi^+}$ as shown in the legend. The errors shown are
  statistical only. See Refs.~\cite{Hakobyan08,Brooks08} for more
  discussion of the preliminary results for positive pions, and
  Ref.~\cite{Hicks09} for preliminary results for neutral kaons.
\label{fig:CLAS_delta_pt2}}
\end{center}
\end{figure}

\subsection{Quark energy loss}
Quarks passing through a strongly-interacting medium are expected to
lose energy through the medium-stimulated emission of gluons. This
topic underwent intensive theoretical study by a number of groups beginning
in the 1990's (see \cite{Baier2000,Accardi2009}). Experimentally this topic was a focus of the Fermilab
Drell-Yan experiments~\cite{Leitch2000} as well as a phenomenon invoked to explain the
jet quenching observed at RHIC. Extraction of a reliable value for the
quark energy loss has proven to be an elusive goal, and estimated
values from both theoretical and experimental work have varied by more
than one order of magnitude.

While obtaining a quantitative value for quark energy loss is still a
challenge, there are closely related parameters that are now
accessible through experiment. Principal among these is the {\em{transport
coefficient}} $\hat{q}$, which quantifies the transverse
momentum accumulated by the quark per unit path length. This parameter
or its equivalent
appears in every calculation of quark energy loss for cold and hot
matter and thus measuring it is of high interest. Using the BDMPS
formalism~\cite{BDMPS97,BDMPS98} for the following discussion, the
transport coefficient is related to the mean transverse
momentum acquired by quark $q$ along a trajectory of length $L$:
\begin{equation}
< p_{T,q}^2 > = \hat{q}L,
\label{eq:q_hat}
\end{equation}
\noindent and the radiative energy loss is given by
\begin{equation}
-\frac{dE}{dz} = {{1}\over{4}}\alpha_s N_c \hat{q} L,
\label{eq:dedz}
\end{equation}
where $\alpha_s$ is the strong coupling constant and $N_c$ is the
number of colors. These equations, while specific to one theoretical
approach, demonstrate the close connection
between energy loss, $p_T$ broadening, and the transport coefficient.

As with the quark energy loss, there is an order of magnitude spread
among the theoretical and experimental estimates for
$\hat{q}$. For example, an estimate from pQCD is the following:

\begin{equation}
  \label{eq:qhatcold}
  \hat{q}\ =\ \frac{4\pi^2\, \alpha_s\ C_R}{N_c^2\ -1}\,
    \rho\, xG(x,Q^2),
\end{equation}
where $C_R$ is the SU(3) color charge of the parton, equal to
$(N_c^2-1)/2N_c$ and $N_c$ for quarks and gluons respectively, $x$ is
the Bjorken scaling variable, $\rho$ is the nuclear medium density,
and $G$ is the gluon density in the nucleon.
The connection between gluon density and $\hat{q}$ is a consistent
theme among the various theoretical approaches.

The new precision Jefferson Lab and HERMES data from the
past decade will provide a definitive measurement of unprecedented
precision for the value of $\hat{q}$ in cold nuclear matter. In this
case, the experimentally measured hadron observable $\Delta p_{T,h}^2$
is very closely related to $\hat{q}$ and rather minor assumptions
are required to derive it from the data.

A related exercise is the extraction of the lifetime of the free
quark, referred to as the `production time' $\tau_p$, from the
data. Although heuristic estimates of this quantity can be found, it
has never been measured experimentally. While an exact form is not
known, its general dependence on $\nu$ and $z$ is expected to be of
the form

\begin{equation}
\tau_p \approx z(1-z)\nu ,
\label{eq:tau_p}
\end{equation}

\noindent based on energy conservation and special
relativity.

The Jefferson Lab data suggest that a unique window exists for measuring
$\tau_p$ where the production length $\tau_p /c$ is comparable to nuclear
dimensions. In this regime, interactions with the nucleus allow a
determination of $\tau_p$, while at higher $\nu$ the quark hadronizes
outside the nucleus and no information on its lifetime can be
gained. The method of extraction makes use of the $p_T$ broadening,
which occurs during the phase in which the quark is propagating as a
colored object. Once it undergoes color neutralization and becomes a
prehadron, its interactions with the nucleus contribute much less
broadening per unit path length. Thus, when the quark's color is
neutralized, the broadening essentially stops. This is manifested
experimentally by a non-linear saturation in the relationship between
$\Delta p_T$ and the nuclear size $\approx A^{1\over{3}}$.

Figure~\ref{fig:CLAS_delta_pt2} shows a small subset of the
preliminary Jefferson Lab 5~GeV data for three different
three-dimensional bins in $Q^2$, $\nu$, and $z$.
On the vertical axis is the experimentally
observed $p_T$ broadening, $\Delta p_{T,h}^2$. On the horizontal axis is
$A^{1\over{3}}$ which is proportional to the average medium thickness
for the three nuclei shown. It is seen that these data exhibit the
non-linear saturation described above, in varying degrees. To
interpret these in terms of quark degrees of freedom one must
correct the hadron's $\Delta p_{T,h}^2$ to the value of the
quark-level $\Delta p_{T,q}^2$, typically using a simple kinematic
multiplier. If the prescription of  Domdey {\it et al.}~\cite{Domdey} is used for this, one
finds that the data shown in this figure follow the behavior expected
from Eq.~(\ref{eq:tau_p}) above. A detailed study of the full data
set is underway, but it is already evident that there is qualitative
confirmation of the physical picture and a clear path to extracting
the free quark lifetime $\tau_p$ from these data.

The determination of $\tau_p$ provides a precise value of $L$ in
Eq.~(\ref{eq:q_hat}), and the value of the quark-level $\Delta p_{T,q}^2$
gives the quantity on the left-hand side of this equation, thus
determining $\hat{q}$. This can then be used to estimate the quark
energy loss through Eq.~(\ref{eq:dedz}).

In principle, the quark energy loss can also be determined from a
direct measurement using the same data and comparing the energy
spectra of leading hadrons emerging from nuclei of different
thicknesses. Careful consideration of Fermi motion and nuclear pion
production are required to extract a direct estimate. An evaluation of
these possibilities is also underway.

\section{Summary}

In this article we have reviewed the tremendous successes of Jefferson Lab
experiments that aimed at the understanding of in-medium modifications
of hadron properties and quark interactions over the last decade.
Particularly, we have highlighted three experiments: First, the
$^4$He$(\vec{e},e'\vec{p})^3$H polarization transfer measurements
revealed possible medium modifications of the bound proton
electromagnetic form factors. Second, the study of vector meson
photoproduction on nuclei has observed no mass shift of the $\rho$
meson in the nuclear medium and a width consistent with expected
collision broadening. These results put tight constrains on models of
in-medium hadron properties. Finally, in-medium modification of
fragmentation function ratios extracted in Jefferson Lab with high
statistics and accuracy, has opened new experimental possibilities that
connect to the future Jefferson Lab 12~GeV upgrade. The data analysis
of some more recent experiments to study the Coulomb sum rule or the
nuclear EMC effect are underway and exciting results from those are
expected to come out soon.

\ack
Notice: Authored by Jefferson Science Associates, LLC under U.S. DOE 
Contract No. DE-AC05-06OR23177. The U.S. Government retains a 
non-exclusive, paid-up, irrevocable, world-wide license to publish or 
reproduce this manuscript for U.S. Government purposes. We would like to 
thank C.~Djalali and J.~Arrington for useful discussions about this 
work.
S.S. acknowledges support from the U.S. National Science
Foundation, NSF PHY-0856010.
W.K.B. acknowledges support from CONICYT grant number 1080564.

\end{document}